\documentclass[aps,prx,reprint,superscriptaddress,footnote, onecolumn,notitlepage]{revtex4-1}

\usepackage{appendix}

\usepackage{tikz}  
\usetikzlibrary{arrows,shapes,positioning,shadows,backgrounds,fit}

\usepackage{relsize}
\usepackage{graphicx}
\usepackage{amsmath, bbm}
\usepackage{amssymb}
\usepackage{amsthm}
\usepackage{mathrsfs}
\usepackage{xcolor}
\usepackage{cancel}
\usepackage{titlesec}
\usepackage{enumitem}  
\usepackage{mathtools}
\usepackage{braket}

\usepackage[normalem]{ulem}

\titlespacing{\section}{0ex}{2ex}{0.4ex}
\titleformat{name=\section}{\bf}{\thesection.}{.3em}{}

\usepackage{soul}
\usepackage{dsfont}

\def\be{\begin{eqnarray}}
\def\ee{\end{eqnarray}}
\newcommand{\tr}[1]{\text{Tr}\left(#1\right)}
\newcommand{\Tr}[1]{\text{Tr}\left[#1\right]}

\theoremstyle{plain}

\definecolor{myblue}{rgb}{0.2,0.2,0.8}
\definecolor{myblack}{rgb}{0,0,0}
\definecolor{myurl}{rgb}{0.1,0.1,0.4}

\usepackage[colorlinks=true,citecolor=myblue,linkcolor=myblack,urlcolor=myurl]{hyperref}

\begin{document}

\title{Covariant currents and a thermodynamic
 uncertainty relation on curved manifolds}

\date{\today}

\author{Harry J.~D. Miller}
\affiliation{Department of Physics and Astronomy, The University of Manchester, Manchester M13 9PL, UK}

\begin{abstract}
A framework for defining stochastic currents associated with diffusion processes on curved Riemannian manifolds is presented. This is achieved by introducing an overdamped Stratonovich-Langevin equation that remains fully covariant under non-linear transformations of state variables. The approach leads to a covariant extension of the thermodynamic uncertainty relation, describing a trade-off between the total entropy production rate and thermodynamic precision associated with short-time currents in curved spaces and arbitrary coordinate systems. 
\end{abstract}

\maketitle

Non-equilibrium thermodynamics is typically expressed in terms of a naturally flat Euclidean phase space or configuration space \cite{risken1996fokker}. However, there is a need to formulate thermodynamics in more general \textit{curved} spaces, motivated by a number of physical systems such as colloidal particles suspended in a spatially inhomogeneous mediums  \cite{celani2012anomalous,durang2015overdamped}, rotating ferromagnetic particles \cite{raible2004langevin}, and diffusion caused by thermophoresis \cite{polettini2013diffusion,polettini2013generally}, along with biological systems such as protein diffusion in cell membranes \cite{metzler2016non} and cell migration on curved surfaces \cite{lin2020collective}. This is further necessitated by ongoing efforts to establish a fully relativistic  non-equilibrium thermodynamics \cite{herrmann2010diffusion,dunkel2009relativistic,cai2023relativistic}. One of the main obstacles is that the standard Langevin and Fokker-Planck equations used in stochastic thermodynamics are not invariant under arbitrary non-linear transformation of state variables, and so must be adapted to preserve covariance. From a first principles point of view any physical theory should be independent of coordinate choices, including the laws of thermodynamics themselves. This point was originally put forward by Graham in 1977 \cite{Graham1977}, who derived a manifestly covariant form of Fokker-Planck equation for describing general state-dependent diffusion processes by interpreting the inverse diffusion tensor as a Riemann metric for the system degrees of freedom. These equations of motion gave rise to formulations of the laws of thermodynamics from a coordinate-independent perspective. This was later extended to diffusion on general Riemannian manifolds \cite{grabert1980fluctuations,Hanggi1982} and also used to establish a covariant form of the Ito-Langevin equation \cite{Graham1985,polettini2013generally}. Brownian motion on curved manifolds has also been extensively explored in mathematics \cite{van1986brownian,hsu2002stochastic,stroock2000introduction}. 

More modern treatments have turned to developing a covariant \textit{stochastic} thermodynamics, which is relevant for small systems where fluctuations play a significant role in the thermodynamic laws and generation of entropy production \cite{seifert2008stochastic,jarzynski2012equalities,seifert2012stochastic,ciliberto2017experiments}. This has included recent generalisations of the fluctuation theorems applicable to systems with coordinates confined to Riemannian manifolds \cite{Ding2022,cai2024fluctuation}. In this paper we continue this effort by developing a fully covariant derivation of the \textit{thermodynamic uncertainty relation} (TUR) \cite{horowitz2020thermodynamic}. The TUR is an important inequality used to quantify the thermodynamics of precision \cite{seifert2018stochastic}, providing a trade-off between the rate of entropy produced by an irreversible process and the noise-to-signal ratio of any stochastic current \cite{barato2015thermodynamic,gingrich2016dissipation,garrahan2017simple,dechant2018multidimensional,brandner2018thermodynamic,hasegawa2019uncertainty,koyuk2019operationally}. This is a convenient tool for ascertaining estimates of the entropy production from observable data \cite{gingrich2017inferring,manikandan2020inferring}, and has been shown to provide fundamental limits to the performances of steady-state heat engines \cite{pietzonka2018universal} and Brownian clocks \cite{barato2016cost}. By going beyond the standard assumptions that use natural coordinates in a flat space, we show how to consistently define time-integrated stochastic currents in curved spaces using a covariant extension of the overdamped Langevin equation. This approach will be used to generalise the TUR to diffusion processes on general Riemannian manifolds. Structurally the TUR we obtain takes the form
\begin{align}\label{eq:TUR}
    \frac{\dot{S}_{\text{tot}}(t)\Delta j_A(t)}{j^2_A(t)}\geq 1,
\end{align}
where $\dot{S}_{\text{tot}}(t)$ is the total entropy production rate, $j_A(t)$ is the average instantaneous current of some observable $A$ while $\Delta j_A(t)$ is a measure of the current fluctuations. This belongs to the class of short-time TURs that have been previously derived from the standard overdamped Langevin equation \cite{Dechant2018,otsubo2020estimating,Dechant2022,dieball2023direct,kamijima2023thermodynamic}. Crucially each term in~\eqref{eq:TUR} is invariant under reparameterisations of state variables, which  means it is possible to estimate the entropy production from observations of currents in any coordinate system of choice. These results thus demonstrate that thermodynamic precision can be quantified consistently on curved manifolds.

Our approach starts with the covariant Langevin equation with multiplicative noise, which is introduced in Section~\ref{sec:I} along with the resulting covariant Fokker-Planck equation describing state-dependent diffusion on Riemann manifolds. In Section~\ref{sec:II} we derive a formula for the irreversible entropy production rate from the covariant Fokker-Planck equation, establishing the statement of the second of thermodynamics in our theory. In Section~\ref{sec:III} we show how to define time-integrated stochastic currents in the covariant theory, and use this to derive the aforementioned TUR~\eqref{eq:TUR}. Finally, in Section~\ref{sec:IV} we apply our results to two examples, including diffusion on the surface of a sphere and a covariant version of the Ornstein–Uhlenbeck process. 

\section{The covariant Langevin and Fokker-Planck equation}\label{sec:I}

\

To begin we will propose a covariant form of the overdamped Langevin equation, blending together a number of previous approaches from the literature \cite{polettini2013generally,Ding2022,Diosi2023a}. Suppose the system is described by $n$ real coordinates $\vec{x}\in\mathcal{M}$ where $\mathcal{M}$ is an $n$-dimensional real manifold with a metric tensor $g_{\nu\mu}$. These coordinates are assumed to be even under time-reversal symmetry so that we may interpret them as spatial variables in a configuration space. The pair $(\mathcal{M},g)$ corresponds to a Riemannian manifold with Levi-Civita connection 
\begin{align}\label{eq:levi}
    {\Gamma^l}_{\nu\mu}=\frac{1}{2}g^{l\kappa}\bigg[\frac{\partial g_{\kappa\mu}}{\partial x^\nu}+\frac{\partial g_{\kappa\nu}}{\partial x^\mu}-\frac{\partial g_{\nu\mu}}{\partial x^\kappa}\bigg].
\end{align}
The choice of the metric should reflect the physical symmetries of the system, though for now it may be left unspecified until we focus on specific examples later in Section~\ref{sec:IV}.

For a deterministic system the equations of motion on $(\mathcal{M},g)$ can generally be represented in a covariant Onsager form $dx^\nu(t)=-L^{\nu\mu}\nabla_\mu \phi \ dt$, where $\phi(\vec{x})$ represents some scalar potential, $\nabla_\mu$ denotes the covariant derivative with respect to connection~\eqref{eq:levi} and $L^{\nu\mu}=L^{\nu\mu}(\vec{x})$ is a a rank-$2$ contravariant tensor of kinetic coefficients \cite{Grabert1980}. The latter may depend on the system variables in which case the equations of motion are non-linear in general. We decompose $L^{\nu\mu}$ into a symmetric and anti-symmetric contribution denoted respectively by
\begin{align}
    D^{\nu\mu}=\frac{1}{2}(L^{\nu\mu}+L^{\mu\nu}), \ \ \ \ \ Q^{\nu\mu}=\frac{1}{2}(L^{\nu\mu}-L^{\mu\nu}).
\end{align}
If the system is also influenced by noise, then the symmetric part $D^{\nu\mu}$ can be used to define a  diffusion tensor for the dynamics. Consistency with the second law requires it to be positive definite, and for technical reasons we also assume it to be non-singular.  We can construct a general  covariant stochastic dynamics on $(\mathcal{M},g)$ using the following non-linear, overdamped Langevin equation with multiplicative noise:
\begin{align}\label{eq:langevinStrat}
    dx^\nu(t)=u^\nu(t)dt+\sigma^\nu_j\circ dW^j(t),
\end{align}
where $\circ$ indicates the Stratonovich product, and $d\vec{W}(t)$ are the set of increments for an $m$-dimensional Wiener process with vanishing mean $\langle d\vec{W}(t) \rangle=0$ and correlations $\langle dW^j(t) dW^k(t')\rangle=\delta^{jk}\delta(t-t')dt$. Here we assume $m\geq n$ to reflect the fact that there are typically more fast environmental variables compared with the slow variables $\vec{x}(t)$, and repeated upper and lower indices in~\eqref{eq:langevinStrat} are implicitly summed over. The drift $\vec{u}(t)$ is a vector field  with contravariant components
\begin{align}\label{eq:drift}
    u^\nu(t):=-L^{\nu\mu}\nabla_\mu \phi+\nabla_\mu Q^{\nu\mu},
\end{align}
The additional divergence term $\nabla_\mu Q^{\nu\mu}$ in the drift may always be included as a sourceless contribution to the covariant Fokker-Planck equation \cite{Graham1977}, and may be viewed as the asymmetric part of the so-called spurious drift \cite{Ding2020}. The noise coefficients are related to the diffusion tensor according to
\begin{align}\label{eq:ortho}
    D^{\nu\mu}=\frac{1}{2}\sigma_j^\nu\sigma_k^\mu\delta^{jk}.
\end{align}
and each $\sigma_j^\nu$ transforms as a contravariant vector component at each fixed index $j$. To see that the Langevin equation is covariant, recall that the contravariant components of a vector change under a non-linear transformation of variables $\vec{x}\mapsto \vec{y}(\vec{x})$ according to the rule
\begin{align}\label{eq:cont}
    A^\nu(\vec{x},t)\mapsto A^{\nu }(\vec{y},t)=\frac{\partial y^\nu}{\partial x^\mu}A^\mu(\vec{x},t)
\end{align}
while covariant components change according to
\begin{align}\label{eq:cov}
    A_\nu(\vec{x},t)\mapsto A_\nu'(\vec{y},t)=\frac{\partial x^\mu}{\partial y^\nu}A_\mu(\vec{x},t)
\end{align}
We will take the usual convention and denote contravariant and covariant vector components  with respective upper and lower indices and keep in mind the relation between them, $A_\nu=g_{\nu\mu} A^\mu$. The contravariance of solution $x^\nu(t)$ of~\eqref{eq:langevinStrat} follows from the fact that the drift~\eqref{eq:drift} is evidently contravariant while the Stratonovich product preserves the chain rule from standard calculus \cite{Graham1985}.

As recently shown by Diosi \cite{Diosi2023a}, one must add an additional covariant constraint on the noise coefficients in~\eqref{eq:langevinStrat} to make sure the Langevin equation is consistent with the corresponding covariant Fokker-Planck equation. The constraint is that the noise coefficients are divergenceless:
\begin{align}\label{eq:constraint}
    \nabla_\nu \sigma^\nu_j=0, \ \ \ j=1,2,..,m.
\end{align}
Combining this with~\eqref{eq:ortho}, one may derive the following identity 
\begin{align}\label{eq:diosi}
    \frac{1}{2}(\partial_\mu\sigma_j^\nu)\sigma_k^\mu\delta^{jk}=\partial_\mu D^{\nu\mu}+\frac{1}{2}\sigma_j^\nu\sigma^\kappa_j\Gamma^\mu_{\mu\kappa}=\partial_\mu D^{\nu\mu}+D^{\nu\kappa}\Gamma^\mu_{\mu\kappa}=\frac{1}{\sqrt{|g|}}\partial_\mu\big(\sqrt{|g|}D^{\nu\mu}\big),
\end{align}
The above identity was used in \cite{Diosi2023a} to ensure consistency with the original covariant Fokker-Planck equation of Graham \cite{Graham1977} using the inverse diffusion tensor as a metric, though it should be stressed that in the present context our equation is more general since one may use any choice of metric. To derive the Fokker-Planck equation we first need to convert our Stratonovich Langevin equation~\eqref{eq:langevinStrat} into an equivalent Ito form. To do so we expand the Stratonovich product in~\eqref{eq:langevinStrat} and combining this with the identity~\eqref{eq:diosi} alongside the definition of the drift~\eqref{eq:drift} we get
\begin{align}\label{eq:langevinIto}
    dx^\nu(t)=\frac{1}{\sqrt{|g|}}\partial_\mu\big(\sqrt{|g|}L^{\nu\mu}\big)dt-L^{\nu\mu}\nabla \phi \ dt+\sigma^\nu_jdW^j(t).
\end{align}
where we made use of the anti-symmetry of $Q^{\nu\mu}$. The product $\sigma^\nu_jdW^j(t)$ should now be understood in the sense of Ito. This is precisely the same form of non-linear, coordinate-invariant Langevin equation proposed by Ding \textit{et al} in \cite{Ding2020}, though they did not impose the constraint~\eqref{eq:constraint}. Without it, the Ito form of~\eqref{eq:langevinIto} does not preserve the contravariance of $x^\nu(t)$ in the sense of first order geometry \cite{cai2024fluctuation}. However, one should also bear in mind that orthogonal transformations of the noise coefficients of form $\sigma^\nu_j\mapsto U^{k}_j(\vec{x})\sigma_k^\nu(\vec{x})$, where $U\in SO(m)$, produce another Langevin equation that is statistically indistinguishable from the original covariant form~\eqref{eq:langevinStrat}  \cite{polettini2013generally}.

The covariant Fokker-Planck equation results from averaging the Langevin equation~\eqref{eq:langevinStrat} over the noise, producing a continuous, \textit{scalar} probability density $\varrho_t(\vec{x})=\langle \tilde{\delta}(\vec{x}(t)-\vec{x}) \rangle$ at time $t$ where $\tilde{\delta}(\vec{x})$ is the scalar Dirac delta function on $(\mathcal{M},g)$ \footnote{The scalar Dirac delta function on a Riemannian manifold $(\mathcal{M},g)$ is defined by $\tilde{\delta}(\vec{x})=\delta(\vec{x})/\sqrt{|g|}$ }. To ensure the correct normalisation condition for $\varrho_t$, we will need the coordinate-invariant volume form of $(\mathcal{M},g)$, 
\begin{align}
    d\Omega=\sqrt{|g|} \ d^n x.
\end{align}
Then the density remains normalised at all times according to
\begin{align}
    \int_{\mathcal{M}}d\Omega \ \varrho_t(\vec{x})=1.
\end{align}
We assume that a steady-state of the dynamics $\pi(\vec{x})$ is determined by the scalar potential, so that
\begin{align}\label{eq:steady}
    \pi(\vec{x})=e^{-\phi(\vec{x})},
\end{align}
is a fixed point of the equations of motion. To derive the equation of motion from the Langevin equation it is more convenient to work with the Ito form~\eqref{eq:langevinIto}, in which case one finds \cite{Ding2020,Ding2022},
\begin{align}\label{eq:FP}
    \dot{\varrho}_t=\frac{1}{\sqrt{|g|}}\partial_\nu\sqrt{|g|}L^{\nu\mu}\bigg(\partial_\mu+(\partial_\mu \phi)\bigg)\varrho_t,
\end{align}
It is also useful to introduce the local mean velocity $v^\nu(t)$, which is a contravariant vector field describing the local average flows through the system and defined 
\begin{align}\label{eq:vel}
    v^\nu(t):=u^\nu(t)-D^{\nu\mu}\nabla_\mu\text{ln} \ \varrho_t,
\end{align}
Since both $\phi$ and $\varrho_t$ are scalar quantities, we can rewrite~\eqref{eq:FP} as a divergence,
\begin{align}\label{eq:FP2}
    \dot{\varrho}_t=-\nabla_\nu v^\nu(t)\varrho_t.
\end{align}
The covariant form of~\eqref{eq:FP2} is now clearly evident, and one can verify that~\eqref{eq:steady} is indeed a steady-state. Together~\eqref{eq:langevinStrat} and~\eqref{eq:FP2} form the basis for investigating the thermodynamics of precision in a fully covariant theory. 

\section{Irreversible entropy production}\label{sec:II}

\

The most fundamental quantity in stochastic thermodynamics is the entropy production which quantifies the amount of irreversibility produced during a non-equilibrium process \cite{seifert2012stochastic}. With the second law of thermodynamics in mind, we can identify it by looking for a positive contribution to rate of change in system entropy. In the covariant theory, the appropriate way to assign entropy is through the differential Shannon entropy of the state at time $t$ on the manifold $(\mathcal{M},g)$, given by
\begin{align}\label{eq:ent}
    S(t):=-\int_{\mathcal{M}} d\Omega \ \varrho_t \ \text{ln} \ \varrho_t,
\end{align}
Unlike the standard differential entropy defined with a flat integration measure $d^n x$, $S(t)$ is invariant under reparameterisation. In fact, we may interpret~\eqref{eq:ent} as a relative entropy with respect to a reference prior given by the volume form $d\Omega=\sqrt{|g|}d^n x$ \cite{polettini2013generally}. The choice of metric thus defines the  maximum entropy state (or state of maximum ignorance) along with the thermostatics of the system \cite{polettini2012nonequilibrium}. When considering the thermodynamics of a non-equilibrium process, we can always consistently decompose the change in system entropy $S(t)$ into a positive entropy production rate together with an environmental contribution, as we now show. 

The key ingredient to determine the entropy production rate is the Riemannian version of Stoke's divergence theorem \cite{carroll2019spacetime}, which states that for a vector field of the form $\vec{F}(\vec{x})\varrho_t(\vec{x})$ the integral of its divergence on manifold $\mathcal{M}$ is equal to the following surface integral over the manifold boundary $\partial \mathcal{M}$:
\begin{align}\label{eq:div}
    \int_{\mathcal{M}} d\Omega \ \nabla_\nu [F^\nu \varrho_t]=\int_{\partial \mathcal{M}} d^{n-1} x \sqrt{|h|} \ \varrho_t \ F^\nu n_\nu
\end{align}
where $\vec{n}$ is the unit normal vector field pointing away from the boundary surface and $h_{\nu\mu}$ represents the metric on $\partial\mathcal{M}$ induced from the pullback of $\mathcal{M}$. To use this, we will assume that the probability density $\varrho_t(\vec{x})$ always decays sufficiently rapidly towards the boundary such that terms arising from the RHS of~\eqref{eq:div} are negligible \cite{VandenBroeck2010}. This assumption may be further extended to manifolds without boundary as a limiting case. Taking the time derivative of the entropy~\eqref{eq:ent} and combining this with the Fokker-Planck equation~\eqref{eq:FP2} gives
\begin{align}\label{eq:ent1}
    \dot{S}(t)&=-\int_{\mathcal{M}} d\Omega \ \dot{\varrho}_t \ \text{ln} \ \varrho_t=\int_{\mathcal{M}} d\Omega \ [\nabla_\nu v^\nu(t)\varrho_t] \ \text{ln} \ \varrho_t=-\int_{\mathcal{M}} d\Omega \   v^\nu(t)\nabla_\nu\varrho_t,
\end{align}
where~\eqref{eq:div} was used in the final equality under the assumption of negligible boundary contributions. Rearranging~\eqref{eq:vel} gives $\nabla_\nu \varrho_t=\varrho_t D_{\nu\mu} \big(u^\mu-v^\mu(t)\big)$ where $D_{\nu\mu}$ denotes the inverse of the diffusion tensor, and then substituting this into the entropy rate~\eqref{eq:ent1} leads to
\begin{align}\label{eq:ent2}
    \dot{S}(t)=\int_{\mathcal{M}} d\Omega \ \varrho_t v^\nu(t)  D_{\nu\mu}v^\mu(t)-\int_{\mathcal{M}} d\Omega \ \varrho_t \  v^\nu(t)  D_{\nu\mu} u^\mu(t).
\end{align}
It is pertinent to introduce the following induced inner product on $\mathcal{M}$, defined by
\begin{align}\label{eq:inner}
    \langle\langle \vec{A},\vec{B} \rangle\rangle_\varrho:=\int_{\mathcal{M}}d\Omega \ \varrho \  A^\nu D^\mu_\nu B_\mu,
\end{align}
where we have used Einstein notation $g^{\mu\kappa}D_{\kappa\nu}=D^\mu_\nu$. The validity of the scalar product is ensured by the positivity and symmetry of the diffusion tensor. Returning to~\eqref{eq:ent2} we identify the following geometric decomposition,
\begin{align}
    \dot{S}(t)=\dot{S}_{\text{tot}}(t)+\dot{S}_{e}(t)
\end{align}
where
\begin{align}\label{eq:irrev}
    \dot{S}_{\text{tot}}(t)=\big\langle\big\langle\vec{v}(t),   \vec{v}(t) \big\rangle\big\rangle_{\varrho_t}\geq 0
\end{align}
is the non-negative total irreversible entropy production rate, and
\begin{align}\label{eq:excess}
    \dot{S}_{e}(t):=-\big\langle\big\langle\vec{v}(t),   \vec{u}(t) \big\rangle\big\rangle_{\varrho_t}
\end{align}
is the entropy flow. The latter has no definite sign and can be interpreted as the entropy flowing into the system from the environment \cite{VandenBroeck2010}. On the other hand, the irreversible contribution $\dot{S}_{\text{tot}}(t)$ is always non-negative and provides a general covariant statement of the second law on $(\mathcal{M},g)$. This generalises the geometric formula for entropy production given in \cite{Dechant2022}, and we recover that result whenever the metric takes a Euclidean form $g^{\nu\mu}\propto \delta^{\nu\mu}$ with constant diffusion matrix $D^\mu_\nu=D\delta^\mu_{\nu}$. 

It is important to note that the total irreversible entropy production accounts for both adiabatic and non-adiabatic contributions \cite{VandenBroeck2010,spinney2012entropy}, and thus can remain non-zero when the system reaches a steady-state. In the present formalism, a necessary condition to maintain a steady-state current is that there must be non-zero anti-symmetric contributions to the kinetic coefficients $L^{\nu\mu}$. To see this we look at the steady-state local mean velocity,
\begin{align}
    v^\nu_{st}=\nabla_\mu Q^{\nu\mu}-Q^{\nu\mu}\nabla_\mu \phi,
\end{align}
which vanishes if $L^{\nu\mu}=D^{\nu\mu}$. If $v^\nu_{st}=0$ the steady-state $\pi(\vec{x})$ is an equilibrium state. In that case the excess entropy rate takes on a more illuminating form, 
\begin{align}
    \dot{S}_e(t):=\int_{\mathcal{M}} d\Omega \ \varrho_t v^\nu(t) \nabla_\nu \phi,
\end{align}
which is simply a sum over the average fluxes associated with the scalar potential.

\section{Covariant currents and thermodynamic uncertainty relation}\label{sec:III}

\

The entropy production~\eqref{eq:irrev} is defined at the ensemble level with respect to the probability density $\varrho_t$. However, we wish to go further and investigate the role of higher order fluctuations in the covariant theory, which means returning to the Langevin equation~\eqref{eq:langevinStrat}. Our core consideration will be a general time-integrated stochastic current \cite{dechant2018current,dieball2022coarse}, defined as the following trajectory dependent and \textit{scalar} quantity 
\begin{align}\label{eq:empcurrent}
    J_{\vec{A}}(\vec{x}(\tau),\tau)=\int^{t=\tau}_{t=0}   A_\nu[\vec{x}(t),t]\circ dx^\nu(t)
\end{align}
where $A_\nu[\vec{x}(t),t]$ is an arbitrary covariant vector field. To prove that the general current~\eqref{eq:empcurrent} is a scalar quantity, we again use the fact that the Stratonovich differential obeys the chain rule, so combining with~\eqref{eq:cov} we have
\begin{align}
    J_{\vec{A}}(\vec{x}(\tau),\tau)\mapsto J_{\vec{A}'}(\vec{y},\tau)=\int^{t=\tau}_{t=0}  A_\nu'[\vec{y}(t),t]\circ dy^\nu(t)=\int^{t=\tau}_{t=0}  \frac{\partial x^\mu}{\partial y^\nu}A_\mu(\vec{x},t) \frac{\partial y^\nu}{\partial x^\kappa}\circ dx^\kappa(t)=J_{\vec{A}}(\vec{x},\tau)
\end{align}
as desired. Therefore, we can safely compute the statistical cumulants of~\eqref{eq:empcurrent} in any coordinate system by solving~\eqref{eq:langevinStrat}. We will now use this current to construct a measure of thermodynamic precision in the covariant theory, which requires one to compute both average rate of change in $J_{\vec{A}}(\vec{x}(\tau),\tau)$ as well as its variance. The precision  measures how reliably the current obtains its average value upon each realisation of the process \cite{otsubo2020estimating,dieball2023direct,kamijima2023thermodynamic}, and is defined as the following signal-to-noise ratio:
\begin{align}\label{eq:precision}
\mathcal{V}_A(t):=\frac{j^2_A(t)}{\Delta j_A(t)},
\end{align}
where
\begin{align}
    j_A(t):=\frac{d}{dt}\langle J_{\vec{A}}(\vec{x}(t),t)\rangle
\end{align}
is the average rate of change in current~\eqref{eq:empcurrent}, while
\begin{align}
    \Delta j_A(t):=\lim_{\tau\to 0} \frac{\text{Var}[J_{\vec{A}}(\vec{x}(t+\tau),t+\tau)]-\text{Var}[J_{\vec{A}}(\vec{x}(t),t)]}{2\tau}
\end{align}
is a measure of the instantaneous change in fluctuations. With these definitions we will now derive the covariant TUR~\eqref{eq:TUR}.

It will convenient to unpack the differential terms in~\eqref{eq:empcurrent}. A key feature of stochastic calculus is the square root scaling of the noise with respect to small increments of time, $d\vec{W}(t)\propto \sqrt{dt}$, meaning that in the continuum limit one has
\begin{align}\label{eq:Ito}
    dx^\nu(t) dx^\mu(t)=\frac{1}{2}D^{\nu\mu} dt,
\end{align}
which holds at the level of the stochastic variables \cite{Hanggi1982}. We use this to expand out the Stratonovich differential in~\eqref{eq:empcurrent}, giving
\begin{align}\label{eq:current1}
    \nonumber J_{\vec{A}}(\vec{x}(\tau),\tau)&=\int^{t=\tau}_{t=0}  A_\nu(t) dx^\nu(t)+\frac{1}{2}\int^{t=\tau}_{t=0}  \partial_\mu A_\nu(t) dx^\mu(t) dx^\nu(t), \\
    \nonumber&=\int^{t=\tau}_{t=0}  A_\nu(t) dx^\nu(t)+\int^{\tau}_{0} dt \   D^{\nu\mu}\partial_\mu A_\nu(t), \\
&=\int^\tau_0 dt \ \bigg[D^{\nu\mu}\partial_\mu A_\nu(t)+\bigg(\frac{1}{\sqrt{|g|}}\partial_\mu (\sqrt{|g|}L^{\nu\mu})-L^{\nu\mu} \partial_\mu \phi \bigg)A_\nu(t)\bigg] +\int^{t=\tau}_{t=0} A_\nu(t)\sigma_j^\nu dW^j(t),
\end{align}
where we used~\eqref{eq:Ito} in the second line and substituted in the Ito form~\eqref{eq:langevinIto} into the last line. While this expression involves partial derivatives, it is more helpful to convert it into an equation involving only the covariant derivatives. Recall that the covariant derivative of covariant vector components is given by
\begin{align}\label{eq:covderiv}
    \nabla_\mu A_{\nu}=\partial_\mu A_\nu-\Gamma_{\mu\nu}^\kappa A_\kappa
\end{align}
The covariant derivative of the symmetric diffusion tensor is
\begin{align}\label{eq:covD}
    \nabla_\mu D^{\nu\mu}=\partial_\mu D^{\nu\mu}+\Gamma^\nu_{\mu\kappa}D^{\mu \kappa}+\Gamma^\mu_{\mu\kappa}D^{\nu\kappa}.
\end{align}
Since $Q^{\nu\mu}$ is anti-symmetric, its divergence is given by the identity
\begin{align}\label{eq:anti}
    \nabla_\mu Q^{\nu\mu}=\frac{1}{\sqrt{|g|}}\partial_\mu (\sqrt{|g|}Q^{\nu\mu}).
\end{align}
Then using the splitting $L^{\nu\mu}=D^{\nu\mu}+Q^{\nu\mu}$ one can combine~\eqref{eq:anti} and~\eqref{eq:covD} to get the identity
\begin{align}\label{eq:trick}
    \frac{1}{\sqrt{|g|}}\partial_\mu (\sqrt{|g|}L^{\nu\mu})=\nabla_\mu Q^{\nu\mu}+\partial_\mu D^{\nu\mu}+\Gamma_{\kappa\mu}^\kappa D^{\nu \mu}=\nabla_\mu L^{\nu\mu}-\Gamma^\nu_{\mu\kappa}D^{\mu\kappa},
\end{align}
where we used $\Gamma^\kappa_{\kappa\mu}=\partial_\mu\text{ln} \sqrt{|g|} $. Combining~\eqref{eq:covderiv},~\eqref{eq:trick} with~\eqref{eq:current1} and noting the scalar property $\partial_\mu\phi=\nabla_\mu \phi$ yields the desired expression for the time-integrated current,
\begin{align}\label{eq:current2}
    J_{\vec{A}}(\vec{x}(\tau),\tau)=\int^\tau_0 dt \ \bigg[D^{\nu\mu}\bigg(\nabla_\mu A_\nu(t)-A_\nu(t)\nabla_\mu\phi\bigg)+\bigg(\nabla_\mu L^{\nu\mu}-Q^{\nu\mu} \nabla_\mu \phi \bigg)A_\nu(t)\bigg]+\int^{t=\tau}_{t=0} A_\nu(t)\sigma_j^\nu dW^j(t).
\end{align}
Taking the time derivative of the average and using the zero-mean property of the Wiener process leads us to
\begin{align}
    \nonumber j_A(t):=\frac{d}{dt}\langle J_{\vec{A}}(\vec{x}(t),t)\rangle &=\int_{\mathcal{M}}d\Omega \  \varrho_t\bigg[\nabla_\mu \big(D^{\nu\mu} A_\nu(t)\big)- D^{\nu\mu} A_\nu(t)\nabla_\mu\phi+\bigg(\nabla_\mu Q^{\nu\mu}-Q^{\nu\mu} \nabla_\mu \phi \bigg) A_\nu(t)\bigg], \\
    \nonumber&=-\int_{\mathcal{M}}d\Omega \  D^{\nu\mu}\bigg(A_\nu(t)\nabla_\mu\varrho_t +\varrho_t A_\nu(t)\nabla_\mu\phi\bigg)+\bigg(Q^{\nu\mu} \nabla_\mu \phi-\nabla_\mu Q^{\nu\mu} \bigg) A_\nu(t), \\
    &=\int_{\mathcal{M}}d\Omega \  \varrho_t \ A_\nu(t) v^\nu(t), 
\end{align}
where we used the divergence theorem~\eqref{eq:div} with vanishing boundary conditions in the second line, along with the  the identity $\nabla_\mu \varrho_t=\varrho_t D_{\mu\kappa} \big(u^\kappa(t)-v^\kappa(t)\big)$ and definition of the drift~\eqref{eq:drift} in the third line. Returning to the inner product~\eqref{eq:inner}, we can now rewrite the average current as
\begin{align}
    j_A(t)=\big\langle\big\langle \mathbf{D}\vec{A}(t),   \vec{v}(t) \big\rangle\big\rangle_{\varrho_t}.
\end{align}
where $\mathbf{D}$ represents a matrix representation of the diffusion tensor $D^{\nu\mu}$. Turning now to the variance, it follows directly from squaring~\eqref{eq:current2} that only the last term will contribute, and so using $\langle dW^j(t) dW^k(t')\rangle=\delta^{jk}\delta(t-t')dt$ we immediately get
\begin{align}
    \Delta j_A(t) =\frac{1}{2}\int_{\mathcal{M}}d\Omega \ \varrho_t A_\nu(t)A_{\mu}(t)\sigma_j^\nu \sigma_k^\mu \delta^{jk}=\frac{1}{2}\int_{\mathcal{M}}d\Omega \ \varrho_t A_\nu(t)A_{\mu}(t)D^{\nu\mu}=\big\langle\big\langle \mathbf{D}\vec{A}(t),   \mathbf{D}\vec{A}(t) \big\rangle\big\rangle_{\varrho_t}.
\end{align}
By the Cauchy-Schwarz inequality,
\begin{align}
    \big\langle\big\langle\vec{v}(t),   \vec{v}(t) \big\rangle\big\rangle_{\varrho_t}\big\langle\big\langle \mathbf{D}\vec{A}(t),   \mathbf{D}\vec{A}(t) \big\rangle\big\rangle_{\varrho_t}\geq \big\langle\big\langle \mathbf{D}\vec{A}(t),   \vec{v}(t) \big\rangle\big\rangle_{\varrho_t}^2
\end{align}
We now recognise the squared velocity term on the LHS as the total entropy production rate~\eqref{eq:irrev}, which is evidently lower bounded by the current precision
\begin{align}\label{eq:TUR2}
\dot{S}_{\text{tot}}(t) \geq \mathcal{V}_A(t).
\end{align}
This concludes the derivation of the short-time covariant TUR~\eqref{eq:TUR}. It is saturated by taking a current proportional to the local mean velocity multiplied by the inverse diffusion tensor, $\vec{A}(t)=\alpha \ \mathbf{D}^{-1}\vec{v}(t)$, which can be written in terms of its contravariant components
\begin{align}\label{eq:equality}
    A^\nu(t)\propto-g^{\nu\mu}\nabla_\mu(\phi+\text{ln} \ \varrho_t)+D^\nu_{\mu}(\nabla_\kappa Q^{\mu\kappa}-Q^{\mu\kappa}\nabla_\kappa \phi).
\end{align}
Note that this TUR holds for any finite-time non-equilibrium process described by the covariant Langevin equation~\eqref{eq:langevinStrat}, and does not require any assumptions about detailed balance,  steady-state or linear response conditions.

\section{Applications to specific models}\label{sec:IV}

\

\noindent In this section we apply the TUR to two examples of stochastic processes on Riemannian manifolds. 

\subsection{Diffusion on a sphere}

For a first example we look at a Brownian particle confined to the surface of a sphere with angular coordinates $\vec{x}=(\theta,\varphi)$ \cite{raible2004langevin,gomez2021geometrical,Ding2022,valdes2023fractional}. If we start from the assumption that our system's diffusion tensor has rotational symmetry with non-equilibrium potential $\phi=\phi(\theta)$, then we necessarily must set the diffusion tensor to be  
\begin{align}
    D^{\theta\theta}=D; \ \ \ D^{\varphi\varphi}=D \ \text{cosec}^2 \ \theta, \ \ \ D^{\theta\varphi}=0.
\end{align}
where $D$ is a constant. It is thus natural to choose a metric proportional to the the inverse diffusion tensor with squared line element $ds^2= d\theta^2+\text{sin}^2 \ \theta d\varphi^2$ and the volume form is $d\Omega=\text{sin} \theta d\theta d\varphi$. The only non-zero Christoffel symbols are 
\begin{align}
   \Gamma^\theta_{\varphi\varphi}=-\text{sin} \theta \  \text{cos}\varphi, \ \ \ \ \Gamma^\varphi_{\theta\varphi}=\Gamma^\varphi_{\varphi \theta}=\text{cot} \ \theta.
\end{align}
It is further assumed that there are no asymmetric kinetic coefficients $Q^{\nu\mu}=0$.  To find the covariant Langevin equation, we will need to solve the covariant constraint~\eqref{eq:constraint} to determine the correct form of the noise coefficients. Note that in general we can satisfy the orthogonality condition~\eqref{eq:ortho} by choosing two vectors of form
\begin{align}
    \nonumber&\vec{\sigma}_1=U(\theta,\varphi)\begin{bmatrix}
           \sqrt{2D} \\
           0 \\          
         \end{bmatrix}, \ \ \ \ \ \ \ \vec{\sigma}_2=U(\theta,\varphi)\begin{bmatrix}
           0 \\
           \sqrt{2D} \ \text{cosec} \ \theta \\          
         \end{bmatrix}, \\
\end{align}
where $U(\theta,\varphi)$ is an arbitrary $2\times 2$ orthogonal matrix, which we can generally write as 
\begin{align}
    U(\theta,\varphi)=\left(\begin{array}{cc}
          \text{cos} \  B(\theta,\varphi) & \mp \  \text{sin} \ B(\theta,\varphi)   \\ 
          \text{sin} \ B(\theta,\varphi)  & \pm \  \text{cos} \  B(\theta,\varphi)  
    \end{array}\right)
\end{align}
so
\begin{align}\label{eq:framesphere}
    &\vec{\sigma}_1=\sqrt{2D} \ \begin{bmatrix}
             \text{cos} \  B(\theta,\varphi) \\
           \text{sin} \  B(\theta,\varphi) \\          
         \end{bmatrix}, \ \ \ \ \ \ \ \vec{\sigma}_2=\pm\sqrt{2D} \ \text{cosec} \ \theta\begin{bmatrix}
           -  \text{sin} \  B(\theta,\varphi) \\
           \text{cos} \  B(\theta,\varphi) \\          
         \end{bmatrix}, 
\end{align}
As shown in Appendix~\ref{app:A}, a solution to~\eqref{eq:constraint} is
\begin{align}\label{eq:angle}
      B(\theta,\varphi)=-\text{arctan}\big(\varphi \ \text{sin} \ \theta\big).
\end{align}
 It is worth remarking that the Langevin equation obtained through the present formalism is slightly different than the Langevin equation given in \cite{Ding2022}, which can be recovered by neglecting the rotation $B(\theta,\varphi)=0$. This choice does not satisfy the covariant constraint~\eqref{eq:angle}, which means the coordinates $\theta(t)$ and $\varphi(t)$ do not transform as proper contravariant vector components. However, the statistics and resulting stochastic thermodynamics are gauge invariant with respect to orthogonal transformations of the noise coefficients \cite{polettini2013generally}, hence the predictions of \cite{Ding2022} are statistically indistinguishable from the Langevin equation we have given here. If we consider any arbitrary instantaneous current on the sphere with covariant components $A_\theta(t)$ and $A_\varphi(t)$, then the corresponding time-integrated current can be decomposed as
\begin{align}
J_{\vec{A}}(\theta(\tau),\varphi(\tau),\tau)=J^{(I)}_{\vec{A}}(\theta(\tau),\varphi(\tau),\tau)+J^{(II)}_{\vec{A}}(\theta(\tau),\varphi(\tau),\tau),
\end{align}
with the deterministic term
\begin{align}
J^{(I)}_{\vec{A}}(\theta(\tau),\varphi(\tau),\tau)=D\int^\tau_0 dt \  \bigg(\partial_\theta A_\theta+\text{sin}(\theta)\partial_\varphi A_\varphi(t)+[\text{sin}^2(\theta)\text{cos} \varphi -\partial_\theta\phi]A_\theta(t)\bigg), 
\end{align}
along with a stochastic term involving two uncorrelated noise processes $dW^1(t)$ and $dW^2(t)$,
\begin{align}
\nonumber J^{(II)}_{\vec{A}}(\theta(\tau),\varphi(\tau),\tau)=\sqrt{2 D}\int^{t=\tau}_{t=0}  \ &[A_\theta(t)\text{cos} (B)\mp A_\varphi(t)\text{sin} (B)\text{cosec} (\theta) ]dW^1(t) \\
& \ \ \ \ \ \ \ \ +[A_\theta(t)\text{sin} (B)\pm  A_\varphi(t)\text{cos} (B)\text{cosec} (\theta) ]dW^2(t)
\end{align}
The covariant Fokker-Planck equation is \cite{raible2004langevin,gomez2021geometrical,Ding2022}
\begin{align}\label{eq:Fpsphere}
    \dot{\varrho}_t=D \ \text{cosec} \ \theta \ \partial_
    \theta  \text{sin} \ \theta\big(\partial_\theta \varrho_t+(\partial_\theta\phi) \varrho_t \big)+D \ \text{cosec}^2  \theta \ \partial^2_\varphi\varrho_t.
\end{align}
Analytic solutions to~\eqref{eq:Fpsphere} are not available, but numerical approaches such as those employed in \cite{raible2004langevin,gomez2021geometrical} can be used to simulate the covariant Langevin equation to ascertain the values of currents and precision~\eqref{eq:precision}. With regard to the TUR, we know from~\eqref{eq:equality} that the current that provides a perfect estimate of the entropy production rate is given by  
\begin{align}
\nonumber&A^\theta(t)=\alpha\partial_\theta \phi(\theta)+\alpha\partial_\theta \text{ln} \ \varrho_t(\theta,\varphi), \\
&A^\varphi(t)=\alpha \ \text{cosec}^2 \theta \ \partial_\varphi\text{ln} \ \varrho_t(\theta,\varphi).
\end{align}
where $\alpha\in\mathbb{R}$.

\subsection{Covariant Ornstein–Uhlenbeck process}

\

A second example we consider is analytically solvable. Let's suppose that the system is described by a scalar quadratic potential of form
\begin{align}
    \phi(\vec{x})=\frac{1}{2} S_{\nu\mu}(\vec{x}) x^\nu x^\mu+c.
\end{align}
with $S_{\nu\mu}$ a positive and non-singular  rank-2 tensor and $c$ a constant. The dimension of the system can be arbitrary, while its diffusion tensor is of a diagonal form
\begin{align}\label{eq:diag}
    D^{\nu\mu}(\vec{x})=e^{-\gamma^{(\nu)}(x^\nu)}\delta^{\nu\mu},
\end{align}
where $\gamma^{(\nu)}(x)$ an arbitrary real function. This includes the class of multi-dimensional Kubo and Gomperz diffusion models \cite{horsthemke1975onsager}. In this case if we  choose the metric as the inverse diffusion tensor then we may reinterpret this system in terms of an underlying Ornstein–Uhlenbeck process \cite{san1979class}, provided that the anti-symmetric part of the kinetic coefficients has a vanishing covariant divergence, $\nabla_{\kappa}Q^{\nu\mu}=0, \ \ \forall \kappa,\nu,\mu$, and similarly $\nabla_{\kappa}S^{\nu\mu}=0, \ \ \forall \kappa,\nu,\mu$. To see this, first one may show that the metric $g_{\nu\mu}=D_{\nu\mu}$ from~\eqref{eq:diag} has vanishing curvature \cite{garrido1982class}, meaning it is isometric to a flat Euclidean space. Therefore we may write $x^\nu=f^\nu( \vec{y})$ as functions of some Euclidean coordinate system $\vec{y}$, in which case all covariant derivatives reduce to partial derivatives and the kinetic coefficients $L^{\nu\mu}$ are constant. We can use this to easily find a set of $n$ orthogonal and divergenceless noise coefficients by using the fact that 
\begin{align}\label{eq:ortho_flat}
    \big[\partial_{\kappa} f^\nu(\vec{y})\big]\big[\partial_{\kappa'} f^\mu(\vec{y})\big]\delta^{\kappa\kappa'}=g^{\nu\mu}.
\end{align}
We therefore fix the noise coefficients to be
\begin{align}\label{eq:frame_flat}
    \sigma_\kappa^\nu(\vec{x})=\partial_\kappa f^\nu(\vec{x})=\sqrt{2}e^{-\frac{1}{2}\gamma^{(\nu)}(x^\nu)}\delta^\nu_\kappa.
\end{align}
which are orthogonal due to~\eqref{eq:ortho_flat}, and have a vanishing divergence since in Euclidean coordinates each coefficient becomes a constant vector, $\sigma_\kappa^\nu=\sqrt{2}\delta^\nu_\kappa$. Clearly the coefficients~\eqref{eq:frame_flat} are only determined up to a constant orthogonal transformation, again reflecting the gauge freedom. The covariant Langevin equation is given by
\begin{align}
    dx^\nu(t)=-\big(e^{-\gamma^{(\nu)}(x^\nu)}\delta^{\nu\mu}+Q^{\nu\mu}(\vec{x})\big)S_{\mu\kappa}(\vec{x})x^\kappa dt+\sqrt{2}e^{-\frac{1}{2}\gamma^{(\nu)}(x^\nu)}\delta^\nu_\kappa\circ dW^\kappa(t),
\end{align}
In Euclidean coordinates this reduces to a canonical Ornstein–Uhlenbeck process in Ito form \cite{uhlenbeck1930theory},
\begin{align}\label{eq:OU}
    dy^\nu(t)=-B^\nu_{\mu}y^\mu dt+\sqrt{2} \ \delta^{\nu}_\kappa dW^\kappa(t),
\end{align}
where 
\begin{align}
B^{\nu}_\mu=\big(\delta^{\nu\kappa}+Q^{\nu\kappa}\big)S_{\kappa\mu}
\end{align}
is the friction matrix and $S_{\nu\mu}$ is independent of $\vec{y}$. This can be solved exactly for any time \cite{garrido1982class}, though here we focus on the long time limit $t\to\infty$ where the system reaches a steady-state given by
\begin{align}
    \pi(\vec{y})=\frac{1}{\sqrt{(2 \pi)^n |S|}}\text{exp}\bigg(-\frac{1}{2} S_{\nu\mu} y^\nu y^\mu\bigg),
\end{align}
The steady state velocity is given by
\begin{align}
    v_{st}^\nu=\lim_{t\to\infty} v^\nu(t)=-Q^{\nu\mu}S_{\mu\kappa} y^\kappa
\end{align}
The steady-state entropy production rate for an Ornstein–Uhlenbeck process is well-known \cite{godreche2018characterising}, and can be expressed entirely in terms of the corresponding anti-symmetric matrix of kinetic coefficients, denoted $\mathbf{Q}$, and the  positive, symmetric covariance matrix $\mathbf{S}$,
\begin{align}
\dot{S}^{(\infty)}_{\text{tot}}=\lim_{t\to\infty}\dot{S}_{\text{tot}}(t)=-\tr{\mathbf{Q}^2 \ \mathbf{S}^{-1}}.
\end{align}
As expected, for a symmetric system with $\mathbf{Q}=0$ then there is no entropy production because the system reaches an equilibrium steady-state. However, a non-zero entropy production rate can be maintained provided that there are anti-symmetric contributions to the Onsager matrix. 

Let's now consider the covariant thermodynamic precision associated with monitoring the time-integrated drift, 
\begin{align}
    J_{\vec{u}}(\vec{x}(\tau),\tau)=\int^{t=\tau}_{t=0}   u_\nu[\vec{x}(t),t]\circ dx^\nu(t).
\end{align}
The average of $J_{\vec{u}}(\vec{x}(\tau),\tau)$ gives the entropy flow~\eqref{eq:excess}, $\langle J_{\vec{u}}\rangle=\int^\tau_0 dt \ \dot{S}_{e}(t) $. Since $J_u(\vec{x}(\tau),\tau)$ is a scalar quantity we can find it by switching to Euclidean coordinates. In the long time limit the entropy flow is
\begin{align}
    \dot{S}_{e}^{(\infty)}=\lim_{t\to\infty} \langle\langle \vec{u}(t),\vec{v}(t)\rangle\rangle=\int_{\mathbb{R}} d^n y \ \pi(\vec{y}) \ y_\nu B_{\mu'}^{\nu} Q^{\mu'\mu}S_{\mu\kappa} y^\kappa=\tr{\mathbf{B}\mathbf{Q}}.
\end{align}
where we used $\langle \vec{y}^T \mathbf{A} \vec{y}\rangle=\tr{\mathbf{A}\mathbf{S}}$ due to the Gaussian form of the steady state. As with the entropy production, the average current rate is only maintained by asymmetric terms of the Onsager matrix. For the instantaneous fluctuations we find
\begin{align}
    \Delta j^{(\infty)}_u=\lim_{t\to\infty} \langle\langle \vec{u}(t),\vec{u}(t)\rangle\rangle=\int_{\mathbb{R}} d^n y \ \pi(\vec{y}) \ y_\nu B^{\nu}_\mu B^\mu_{\kappa} y^\kappa=\tr{\mathbf{B}^T\mathbf{S}\mathbf{B}}.
\end{align}
Finally, we can check consistency with the TUR~\eqref{eq:TUR} using 
\begin{align}
    \dot{S}_{\text{tot}}^{(\infty)} \ \Delta j_u^{(\infty)}=\tr{(\mathbf{Q}\mathbf{S}^{-1/2})^T (\mathbf{Q}\mathbf{S}^{-1/2})}\tr{(\mathbf{S}^{1/2}\mathbf{B})^T \mathbf{S}^{1/2}\mathbf{B}}\geq |\tr{\mathbf{B}\mathbf{Q}}|^2=\big(\dot{S}^{(\infty)}_{e}\big)^2.
\end{align}
which in this case follows directly from the Cauchy-Schwarz inequality for the real matrix inner product $\Tr{\mathbf{A}^T \mathbf{B}}$. The condition of equality for the TUR can be stated in matrix terms,
\begin{align}
    \mathbf{Q}\propto\mathbf{S}^{1/2} \mathbf{B} \mathbf{S}^{1/2}, 
\end{align}
We conclude that in this example, the covariant entropy production and thermodynamic precision in the long time limit are ambivalent to the specific form of the diffusion coefficients in~\eqref{eq:diag}, since this entire family of models with non-constant diffusion are isometrically equivalent to a canonical Ornstein–Uhlenbeck process~\eqref{eq:OU}. This provides a clear illustration of the covariant features of the formalism, with the defined entropy production and current precision characterising the intrinsic features of the stochastic thermodynamics independent of the coordinate system. 
 
\section{Concluding remarks}

\

In summary, we have shown how to consistently identify covariant stochastic currents in diffusion processes on curved Riemannian manifolds, and derived a corresponding thermodynamic uncertainty relation~\eqref{eq:TUR2} valid beyond Euclidean configuration spaces. To do so we introduced an overdamped Stratonovich-Langevin~\eqref{eq:langevinStrat} that differs from other approaches in the literature such as \cite{Graham1985,Ding2022}, but has the benefit of being fully covariant at the level of first order geometry without any modification to the Ito differential. The key ingredient was to impose an additional covariant constraint on the noise coefficients, extending the approach taken in \cite{Diosi2023a}. We used this equation to define stochastic currents~\eqref{eq:empcurrent} and their corresponding thermodynamic precision~\eqref{eq:precision} from a coordinate-independent perspective. Ultimately this could be useful for estimating entropy production in more convenient coordinate systems. This should be especially relevant to spatially inhomogeneous systems and biological processes that exhibit diffusion within curved geometries.  

There is scope for further developments of this work. Firstly we have only been concerned with overdamped systems, but underdamped systems could also be explored using a more general covariant Langevin equation. Note that underdamped covariant stochastic dynamics have been considered elsewhere \cite{polettini2013generally,cai2024fluctuation}. Secondly, we have neglected to include any non-conservative forces in the Langevin equation; extending the TUR to such systems could be achieved with the formalism outlined in \cite{ding2022unified}, for example. Furthermore, it is worth emphasising that the results we have obtained all stem from the introduction of the scalar product~\eqref{eq:inner}. For standard flat manifolds this inner product has been used for a number of purposes, such as connecting entropy production to information-geometry \cite{ito2020stochastic}, optimal transport \cite{nakazato2021geometrical,Ito2023a} and alternative geometric decompositions \cite{Dechant2022}. Naturally one could use the present formalism to pursue similar ideas with more general curved manifolds. Finally, we also anticipate that this framework could be relevant to quantum mechanical systems. For example, geometric representations of quantum theory typically involve diffusion on a curved state manifold \cite{hughston1996geometry,brody1999thermalization}, and a covariant thermodynamic approach could provide a useful tool for understanding irreversibility and precision in quantum regimes.

\

\noindent \emph{Acknowledgements:} H.M. acknowledges funding from a Royal Society Research Fellowship (URF/R1/231394)

\bibliography{mybib.bib}

\appendix

\section{Finding divergenceless frame vectors on a sphere }\label{app:A}

\

Taking the metric to be $ds^2= d\theta^2+\text{sin}^2 \ \theta d\varphi^2$, the divergence of any contravariant vector field $V^\nu$ can be found from the general formula
\begin{align}
    \nabla_\nu V^\nu=V^\theta\text{cot} \ \theta +\partial_\theta V^\theta+\partial_\varphi V^\varphi
\end{align}
So plugging in the various components of the two frame vectors in~\eqref{eq:framesphere} we find
\begin{align}
    \nonumber&\nabla_\nu \sigma^\nu_1\propto
\text{cos} \  B \  \big(\text{cot} \  \theta+\partial_\varphi B\big)- \text{sin} \ B\ \partial_\theta B \\
    &\nabla_\nu \sigma^\nu_2\propto -\text{cosec} \theta \  \bigg(\text{sin} \ B \ \partial_\varphi B+\text{cos} \ B \  \partial_\theta B\bigg)
\end{align}
Setting these to zero in accordance with the desired covariant constraint~\eqref{eq:constraint} we find two simultaneous equations
\begin{align}\label{eq:no1}
\partial_\theta B &= \frac{1}{2}\text{sin}  (2B)  \text{cot} \ \theta \\
\partial_\varphi B &=-\text{cos}^2(B) \text{cot} \ \theta, \label{eq:no2}
\end{align}
These equations are solved by
\begin{align}
B(\theta,\varphi)=-\text{arctan}\big(\varphi\text{sin} \ \theta\big).
\end{align}
Note that this solution is only unique up to a constant shift in $\varphi$.

\end{document}